\documentstyle[12pt]{article}

\def\ao{\~ao }

\def\({\c c}
\def\|{\'\i}

\begin{document}

\begin{flushright}   
SLAC-PUB-8591 \\
August 2000
\end{flushright}

\medskip

\baselineskip=18pt  %

\begin{center}
{\large\bf LIGHT-FRONT QCD IN LIGHT-CONE GAUGE}\footnote{Research
partially supported by the Department of Energy under
contract DE-AC03-76SF00515 }

\vspace{1cm}

\baselineskip=20pt  %
{\large Prem P. Srivastava}\footnote{E-mail:\quad prem@uerj.br 
or prem@lafex.cbpf.br } 

\vspace{0.3cm}
{\em\it Instituto de F\'{\i}sica, Universidade do Estado de Rio de Janeiro, 
  Rio de Janeiro, RJ}\\
and \\
{\large Stanley J. Brodsky}\footnote{E-mail:\quad sjbth@slac.stanford.edu}

\vspace{0.3cm}
{\em \it Stanford Linear Accelerator Center, Stanford University,
Stanford, California 94309}
\vspace{0.2cm}
\end{center}

\vfill


\begin{center}
{\bf Abstract}
\end{center}

{\small\quad The light-front (LF) quantization \cite{ref:a} of QCD in 
light-cone (l.c.) gauge is discussed.   
The Dirac method is employed to construct  the LF 
Hamiltonian and theory quantized canonically.  The Dyson-Wick perturbation theory expansion 
based on LF-time ordering is constructed. 
The framework   incorporates in it simultaneously  
 the Lorentz gauge condition as an operator equation as well.  
The propagator of the dynamical 
 $\psi_+$ part of the free fermionic
propagator  is  
 shown  to be causal while the gauge field propagator 
 is found to be transverse.  
  The  interaction Hamiltonian is re-expressed 
 in the form closely resembling the one in covariant theory, except for 
additional  instantaneous interactions, which can be treated systematically.   
Some explicit computations in QCD  are  given. }
\vfill

\begin{center}
Presented at VII Hadron Physics 2000 \\
Caraguatatuba, S\ao Paulo, Brazil \\
10--15 April 2000
\end{center}
\newpage

\baselineskip=17pt     

\section{Introduction}
The quantization of relativistic field theory at fixed light-front time
$\tau = (t - z/c)/\sqrt 2$,  was proposed by Dirac \cite{dir} half a
century ago. It has found important applications \cite{bro,bpp,ken,pre} in
gauge theory and string theory.    The
light-front (LF) quantization of QCD in its Hamiltonian form provides an
alternative approach to lattice gauge theory for the computation of
nonperturbative quantities. 
We discuss here \cite{lcg}  the LF quantization of QCD 
 gauge field theory in l.c. gauge employing the  Dyson-Wick S-matrix 
expansion  based on LF-time-ordered products. The case of covariant 
gauge has been  discussed in our earlier work \cite{cov}.   
\section{ QCD action in  light-cone gauge}
 The LF   coordinates  are defined as 
$x^{\mu}=(x^{+}=x_{-}=(x^{0}+x^{3})/{\sqrt 2},\, x^{-}=x_{+}=
(x^{0}-x^{3})/{\sqrt 2},\, x^{\perp})$, where $x^{\perp}= (x^{1},x^{2})
=(-x_{1},-x_{2})$ 
are the transverse coordinates and $\,\mu=-,+,1,2$.   
The coordinate  $x^{+}\equiv \tau$ will be taken as 
the LF time, while $x^{-}$ is the longitudinal spatial coordinate.   

The quantum action 
 of  QCD in  l.c.  gauge 
is   described in the  standard notation  by 
\begin{equation}
{\cal L}_{QCD}=-{1\over 4}F^{a\mu\nu}{F^{a}}_{\mu\nu}+ B^{a}
{A^{a}}_{-} + {\bar c}^{a}{\cal D}^{ab}_{-} c^{b}
+ {\bar\psi}^{i}
(i\gamma^{\mu}{D^{ij}}_{\mu}-m\delta^{ij})\psi^{j}. \nonumber 
\end{equation}
Here $\,{\bar c}^{a}, c^{a}$ are 
anticommuting   ghost fields and  
  auxiliary   fields $B^{a}(x)$  are introduced in 
the {\it linear} gauge-fixing term.  
  The action is   invariant under 
BRS symmetry  transformations. Since $B^{a}$ carries 
canonical dimension three no  
 quadratic terms in them are permitted.

\section{Spinor field propagator on the LF}
 The quark field term in  LF coordinates reads 
\begin{eqnarray}
{\bar\psi}^{i}
(i\gamma^{\mu}{D^{ij}}_{\mu}-m\delta^{ij})\psi^{j}
&=& i{ \sqrt {2}}{\bar\psi_{+}}^{i}
\gamma^{0}{D^{ij}}_{+}{\psi_{+}}^{j}+{\bar\psi_{+}}^{i}
(i\gamma^{\perp}{D^{ij}}_{\perp}-m\delta^{ij}){\psi_{-}}^{j} \nonumber \\
&+&{\bar\psi_{-}}^{i}\left[{i \sqrt {2}}\gamma^{0}
{D^{ij}}_{-}{\psi_{-}}^{j}+ (i\gamma^{\perp}{D^{ij}}_{\perp}-m\delta^{ij})
{\psi_{+}}^{j}\right]
\end{eqnarray}
where \cite{cov} ${\psi_{\pm}}=\Lambda^{\pm}\psi$. 
This  shows that the  minus components ${\psi_{-}}^{j}$  are in fact nondynamical 
 fields without  kinetic terms. 
Their equations of motion in l.c. gauge  lead to 
the constraint equations  
\begin{equation}
i {\sqrt 2} \,{\psi_{-}}^{j}(x)
= -
\frac{1}{\partial_{-}}\,\,
(i\gamma^{0}\gamma^{\perp}{D^{kl}}_{\perp}-m\gamma^{0}\delta^{kl})
\,\,{\psi_{+}}^{l}(x). \nonumber
\end{equation}
The free  field 
propagator of $\psi_{+}$  is determined from 
the  quadratic terms  (suppressing the 
color index) 
$ i\sqrt {2} {\psi_{+}}^{\dag}\partial_{+}\psi_{+}
+{\psi_{+}}^{\dag}(i\gamma^{0}\gamma^{\perp}\partial_{\perp}- 
m\gamma^{0})\psi_{-}$ 
where 
$\,2i\partial_{-}\psi_{-}$$=(i\gamma^{\perp}\partial_{\perp}+m)
\gamma^{+}\psi_{+}\,$.     The equation of motion for the independent 
component $\psi_{+}$ is nonlocal in the longitudinal direction. In the 
quantized theory we find  
 the following nonvanishing {\it local} anticommutator 
$ \{{\psi}_{+}(\tau, x^{-},x^{\perp})
,{{\psi}_{+}}^{\dag}(\tau, y^{-},y^{\perp})\}$$=
 {1\over \sqrt {2}} 
 \Lambda^{+} \delta(x^{-}-y^{-})\delta^{2}(x^{\perp}-y^{\perp})$. 
They may be realized in momentum space through the following Fourier 
transform \cite{cov} 
\begin {eqnarray}
\psi(x)={1\over {\sqrt {(2\pi)^{3}}}}
              \sum_{r={\pm}}\int d^{2}p^{\perp}
	      dp^{+}&
	      \theta(p^{+})&{\sqrt{m\over p^{+}}}\left[b^{(r)}(p){ u^{(r)}}(p)
	       e^{-ip\cdot x} \right. \nonumber \\
&+&
        \left.  d^{{\dag}{(r)}}(p){ v^{(r)}}(p) e^{ip\cdot x}\right]
\end {eqnarray}  
where 
 \begin{equation}
u^{(r)}(p)
= \frac{1}{({ {\sqrt {2}}p^{+} m })^{1\over 2}}\left[{\sqrt 2} p^{+}\Lambda^{+} 
+(m+\gamma^{\perp}p_{\perp})
\Lambda^{-}\right] \tilde u^{(r)} \nonumber 
\end{equation}
and the
nonvanishing anticommutation relations are 
given by: $\{b^{(r)}(p),{b^{\dag}}^{(s)}(p')\}$
$=\{d^{(r)}(p),{d^{\dag}}^{(s)}(p')\}= \delta_{rs}
\delta(p^{+}-p'^{+})\delta^{2}(p^{\perp}-p'^{\perp})$.

The free propagator then follows to be \cite{cov} 
    \begin{equation}
<0|T({\psi^{i}}_{+}(x){\psi^{\dag j}}_{+}(0))|0>  =   
{{i\delta^{ij}}\over {(2\pi)^{4}}} \int d^{4}q \;{{{\sqrt {2}}q^{+}\, 
\Lambda^{+} }\over 
{(q^2-m^{2}+i\epsilon)}}\, e^{-iq.x}. \nonumber
\end{equation}
It is causal and  contains   no   instantaneous  
 term.  
\section{Gauge field propagator in l.c.  gauge}

 In the l.c. gauge the ghost fields decouple and  it is   sufficient   
to study   the free abelian gauge theory with the 
action  
\begin{equation}
\int d^{2}x^{\perp}dx^{-} \left\{{1\over 2}\left[(F_{+-})^{2}-
(F_{12})^2 +2F_{+\perp}
F_{-\perp}\right]+ B A_{-} \right\}\nonumber
\end{equation}
where    
$F_{\mu\nu}=$  $(\partial_{\mu}A_{\nu}-\partial_{\nu}A_{\mu})$.
Following the Dirac procedure we show that the phase space constraints 
remove all the canonical momenta from the theory. The 
  surviving  variables are 
 $A_{\perp}$ and  $A_{+}$. The latter, however, is a dependent variable 
 satisfying  
   $\partial_{-}(\partial_{-}A_{+}-\partial_{\perp}A_{\perp})=0$.  
The construction of the Dirac bracket shows that in the l.c. gauge on the LF 
we simultaneously obtain  the Lorentz condition $\partial\cdot A=0$ as 
an operator equation as well.  
   The {\it reduced}  Hamiltonian is  found to be 
$
{H_{0}}^{LF} = 
{1\over2}\int {d^{2}x^{\perp}}dx^{-}\; 
\left[(\partial_{-}A_{+})^{2} + \frac{1}{2} F_{\perp\perp'}F^{\perp\perp'}
\right]$

The equal-$\tau$ commutators are
 $[A_{\perp}(x),A_{\perp}(y)]$
$=i\delta_{\perp\perp'} K(x,y)$ 
 where $K(x,y)=-(1/4)\epsilon(x^{-}-y^{-})\delta^{2}(x^{\perp}-y^{\perp})$.  
They  are   {\it nonlocal} in the longitudinal coordinate but there is 
no violation  of the microcausality principle on the LF. 
They  
may be  realized in 
 momentum  space by the following Fourier transform \cite{lcg}
\begin{eqnarray}
A^{\mu}(x)&=&{1\over {\sqrt {(2\pi)^{3}}}}
\int d^{2}k^{\perp}dk^{+}\,
{\theta(k^{+})\over {\sqrt {2k^{+}}}} \nonumber\\
&& \sum_{(\perp)} {E_{(\perp)}}^{\mu}(k)
\left[b_{(\perp)}( k^{+},k^{\perp})
 e^{-i{ k}\cdot{x}}
+b^{\dag}_{(\perp)}(k^{+},k^{\perp})
 e^{i{ k}\cdot{ x}} \right ]. 
\end{eqnarray}
 where $k^{-}$ is shown \cite{lcg} to be defined through 
the dispersion relation, $\, 2 k^{-}k ^{+}= k^{\perp}k^{\perp}$ 
corresponding to a {\it massless} photon. 
Here the nonvanishing commutators are  given by 
$[b_{(\perp)}(k),{b^{\dag}}_{(\perp')}(k')
]$ $=\delta_{(\perp)(\perp')}$
$\delta^{3}(k-k')$. 
The free gluon propagator is hence found to be \cite{lcg}
\begin{eqnarray}
<0|\,T({A^{a}}_{\mu}(x){A^{b}}_{\nu}(0))\,|0> 
&=&{{i\delta^{ab}}
 \over {(2\pi)^{4}}} 
 \int d^{4}k \;e^{-ik\cdot x}  \; \; 
 {D_{\mu\nu}(k)\over {k^{2}+i\epsilon}},  \\
   D_{\mu\nu}(k)&=& D_{\nu\mu}(k)= -g_{\mu\nu} 
+ \frac {n_{\mu}k_{\nu}+n_{\nu}k_{\mu}}{(n\cdot k)}
                - \frac {k^{2}} {(n\cdot k)^{2}} \, n_{\mu}n_{\nu}\nonumber
\end{eqnarray}
with    
 $ n_{\mu}={\delta_{\mu}}^{+}$,  
 ${E_{(\perp)}}^{\mu}(k)=$$ E^{(\perp){\mu}}(k)$$= - D^{\mu}_{\perp} (k)$, 
 $k^{\mu}D_{\mu\nu}(k)=0$, $n^{\mu}D_{\mu\nu}(k)=0$.
%

\section {QCD Hamiltonian in l.c. gauge}

The interaction Hamiltonian in the l.c. gauge, $A^{a}_{-}=0$,  may be 
rewritten \cite{lcg} as   
\begin{eqnarray}                      
{\cal H}_{int}&=& 
  +\frac{g}{2}\,
f^{abc} \,(\partial_{\mu}{A^{a}}_{\nu}-
\partial_{\nu}{A^{a}}_{\mu}) A^{b\mu} A^{c\nu} 
 +\frac {g^2}{4}\, 
f^{abc}f^{ade} {A_{b\mu}} {A^{d\mu}} A_{c\nu} A^{e\nu}\nonumber \\
&& -g \,{{\bar\psi}}^{i}
\gamma^{\mu}{A_{\mu}}^{ij}{{\psi}}^{j}  
 - \frac{g^{2}}{ 2}\,\, {{\bar\psi}}^{i}
\gamma^{+}
\,(\gamma^{\perp'}{A_{\perp'}})^{ij}\,\frac{1}{i\partial_{-}} \, 
(\gamma^{\perp} {A_{\perp}})^{jk}\,{\psi}^{k} \nonumber \\
&& -\frac{g^{2}}{ 2}\, j^{+}_{a}\, \frac {1}{(\partial_{-})^{2}}\, 
j^{+}_{a}
\end{eqnarray}  
where 
$j^{+}_{a}={{\bar\psi}}^{i}
\gamma^{+} ( {t_{a}})^{ij}{{\psi}}^{j}
+ f_{abc} (\partial_{-}
A_{b\mu}) A^{c\mu}$ and 
 a sum over distinct flavours, not written explicitly, 
 is to be understood.  

The fact that gluons have only physical degrees of freedom in l.c. gauge may 
provide an analysis of coupling renormalization similar to that of the pinch 
technique, which is currently being discussed \cite{pinch} in order to obtain a 
shorter expansion and scheme for QCD. In addition, the 
couplings of gluons in the l.c. gauge provides a simple procedure 
for the factorization of soft and 
hard gluonic corrections in high momentum transfer inclusive and exclusive 
reactions.

\end{document}